\documentclass[iop,tighten,apjl]{emulateapj}

\def \aj {AJ}
\def \mnras {MNRAS}
\def \apj {ApJ}
\def \apjs {ApJS}
\def \apjl {ApJL}
\def \aap {A\&A}

\def \araa {ARAA}
\def \pasp {PASP}

\usepackage{amssymb,epsfig,natbib}

\shorttitle{Bars rejuvenating bulges}
\shortauthors{P. Coelho \& D. A. Gadotti}

\begin{document}

\slugcomment{To appear in The Astrophysical Journal Letters}

\title{Bars rejuvenating bulges? Evidence from stellar population analysis}
\author{P. \ Coelho\altaffilmark{1}}
\affil{N\'ucleo de Astrof\'{\i}sica Te\'orica, Universidade Cruzeiro do Sul, 
R. Galv\~ao Bueno 868, Liberdade, 01506-000, S\~ao Paulo, Brasil; paula.coelho@cruzeirodosul.edu.br}

\author{D. A. \ Gadotti\altaffilmark{2}}
\affil{European Southern Observatory, 
Casilla 19001, Santiago 19, Chile; dgadotti@eso.org}

\begin{abstract}
We obtained stellar ages and metallicities via spectrum fitting for a sample of 575 bulges with spectra available from the Sloan Digital Sky Survey.
The structural properties of the galaxies have been studied in detail in \citet{gadotti09} and the sample contains 251 bulges in galaxies with bars.
Using the whole sample, where galaxy stellar mass distributions for barred and unbarred galaxies are similar, we find that bulges in barred and unbarred galaxies occupy similar {\it loci} 
in the age vs.\, metallicity plane. However,
the distribution of bulge ages in barred galaxies shows an excess of populations younger than
$\sim\,$4\,Gyr, when compared to bulges in unbarred galaxies. Kolmogorov-Smirnov statistics confirm
that the age distributions are different with a significance of 99.94\%.  If we select sub-samples for which the {\em bulge} stellar mass distributions are similar for barred and unbarred galaxies, this excess vanishes for galaxies with bulge mass $\log M<10.1\,M_\odot$ while for more massive galaxies we find a bimodal bulge age distribution for barred galaxies only, corresponding to two normal distributions with mean ages of 10.4 and 4.7 Gyr.
We also find twice as much AGN among barred galaxies, as compared to unbarred galaxies, for low-mass bulges.
By combining a large sample of high quality data with sophisticated image and
spectral analysis, we are able to find evidence
that the presence of bars affect the mean stellar ages of bulges.
This lends strong support to models in which bars trigger star formation activity in the centers of galaxies.
\end{abstract}


\section{Introduction}

A large number of studies have investigated the impact of bars on the evolution of galaxies \citep[see reviews by][and references therein]{sellwood93,kormendy04,gadotti_bars}. One of the expectations that emerges from both observation and theory, in the framework of secular evolution processes induced by bars, is the rejuvenation of the stellar population in the central structural component of disk galaxies. Bars are able to collect gas in the disk from within the bar ends to the central parts of the disk, supposedly helping the building of bulges through central star formation episodes \citep[see][]{athanassoula05}.
Simulations such as those of \citet{athmir02}, but including gas, show that the transfer of gas to the center should be fast, $\approx10^8$ yr (E. Athanassoula, priv. comm.).

To date, there is evidence for an enhanced star formation rate, i.e. {\em current} star-forming activity, in the centers of barred galaxies, mostly from studies of nuclear H{\sc ii} regions. For instance, \citet{ho97} found that H$\alpha$ emission line luminosities and equivalent widths are enhanced in barred galaxies, as compared to unbarred galaxies, when one considers early-type disk galaxies only \citep[see also][]{huang96,alonso-herrero01,jogee05,ellison11}.

Direct evidence supporting bulge building by bars from the {\em ages of stars in bulges} has proven to be much more elusive. Studies based on integrated colors have to deal with uncertainties that arise from the age-metallicity degeneracy and effects of dust extinction \citep[see e.g.][]{gadotti01}. On the other hand, studies based on stellar spectral analysis have been, to date, handicapped by poor statistics \citep{peletier07,perez11}. 

In this Letter, we use spectral analysis techniques on a large and well defined sample of barred and unbarred disk galaxies, drawn from SDSS, allowing us to compare mean stellar ages of bulges in barred and unbarred galaxies with unprecedented statistical significance. We describe the most relevant features of our data in the next section. Section \ref{sec:analysis} describes relevant details of our spectral analysis techniques, and results are shown in Sect. \ref{sec:res}. We present our main conclusions in Sect. \ref{sec:con}.

\section{Sample}
\label{sec:sample}

The sample used here is based on the one studied in \cite{gadotti09}\footnote{See \url{http://www.sc.eso.org/~dgadotti/buddaonsdss.html}.}, which contains all
galaxies in SDSS Data Release 2 with stellar masses larger than $10^{10}$\,M$_\odot$ \citep[from][]{kauffmann03}, at redshift
0.02\,$\leq$\,z\,$\leq$\,0.07, and with axial ratio b/a\,$\geq$\,0.9. 
These criteria provide a
sample which is both representative and suitable for 2D bulge/disc/bar
decomposition, as selecting face-on galaxies minimizes dust and
projection effects and eases the identification of bars. The reader is referred to that paper for a detailed discussion on selection effects. Through 2D decomposition, \citet{gadotti09} provides reliable structural parameters, such as the total stellar mass, bulge stellar mass, bulge effective radius $r_e$, bulge/total and bar structural parameters, which are used in this Letter. In short, bulge stellar masses were obtained from bulge luminosity and mass-to-light ratio in the $i$-band, the latter derived from the bulge $g-i$ color. Galaxy total stellar masses were obtained by adding the masses of its components.

To verify whether a galaxy is barred, typical bar signatures were
searched for through inspection of the galaxy image, isophotal
contours and a pixel-by-pixel radial intensity profile 
(as in \citealt{2010MNRAS.407L..41S} and \citealt{2010PASP..122.1397S}). It should be
noted that due to the limited spatial resolution of SDSS images, we
miss most bars with semimajor axis shorter than $\sim2-3$\,kpc, which are
found mainly in very late-type spirals (later than Sc; \citealt{elmegreen2.85}). 

We selected from the previous sample all disk galaxies with bulges and filtered 
them in terms of bulge-to-total and signal-to-noise ratios 
(see \S\ref{sec:res}) so that our effective sample has 251 barred and 324 unbarred galaxies, of which 187 are AGNs
 according to the classification in \citet{kau03}. No galaxy shows
emission lines with equivalent width larger than 20\AA\ and therefore, 
according to the criterium in \citet[][table 1.1]{peterson97}, all AGNs are type 2.

The galaxy stellar mass distribution is similar for barred and unbarred galaxies, with a difference with statistical significance of less than 1$\sigma$, according to a Kolmogorov-Smirnov test (hereafter KS). This means that possible biases, which would be caused by different mass distributions, are absent in our results. For instance, this indicates that global star formation histories are similar in the barred as in the unbarred galaxies in our sample \cite[see][]{kauffmann03}.


The spectra, obtained from the SDSS database, 
have a spectral resolution of $\lambda/\Delta\lambda\sim\,$1800 \citep{york+00}
and an average
S/N in the spectral region covered by the SDSS $g$-band of $\sim\,$21. The $g$-band spectral region is most relevant for our analysis, as it contains the spectral features most sensitive to the stellar parameters we aim at deriving, i.e. ages and metallicities. The spectra were brought to restframe 
and corrected for Galactic extinction as in \citet{cid+05}. It should be noted that SDSS spectra are taken through a fixed fiber size on sky of 3", centered at the galaxy center. \citet{gadotti09} shows that for most galaxies in his sample the light within the fiber diameter is emitted mainly by bulge stars. In this study, we have used the results from the decompositions in \citet{gadotti09} to measure the disk contribution. We verify, through KS tests, if the distributions of such disk contamination inside the fiber are similar for barred and unbarred galaxies, avoiding possible related biases. This is discussed on a case-by-case basis. 

\section{Analysis}

\label{sec:analysis}
\begin{figure}
\begin{center}
\epsfig{file=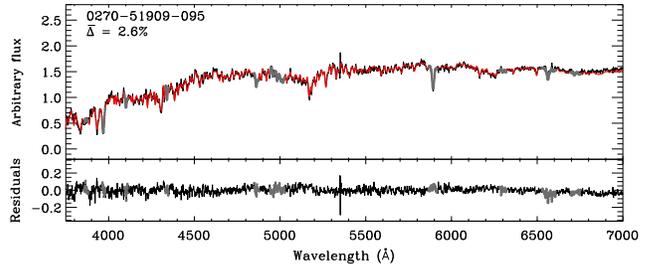,bbllx=30,bblly=345,bburx=505,bbury=550,clip=true,width=\columnwidth}
\caption{Example of a {\sc Starlight} fit. The galaxy identification corresponds to SDSS plate-mjd-fiber numbers.
Observed spectrum and model fit are shown as black and red lines, respectively. 
Pixels not considered
in the spectral fitting (either emission lines or clipped pixels), are marked as gray points. The average absolute deviation 
$\overline\Delta = (\sum{| (M_\lambda-O_\lambda)/O_\lambda|})/N_{pixels}$
of the fit is given. The bottom panel shows the residuals $O_\lambda-M_\lambda$.
}
\label{f:fit}
\end{center}
\end{figure}

We use the spectrum fitting code 
{\sc Starlight}\footnote{See \url{http://www.starlight.ufsc.br}.} \citep{cid+05} 
to compare, on a pixel-by-pixel basis, the SDSS spectra to stellar population (SP) models. 

In short, {\sc Starlight} mixes different computational techniques to fit an observed spectrum $O_\lambda$
with a combination of $N_*$ simple stellar population (SSP) models. Extinction is modelled
as due to foreground dust and 
line-of-sight
stellar motions are modelled by a Gaussian distribution $G$ centered at velocity $v_*$
and with dispersion $\sigma_*$. Both kinematical and SP parameters are derived 
during the fit and the best model spectrum is given by:

\begin{equation}
M_\lambda=M_{\lambda_0} \Bigg( \displaystyle\sum\limits_{j=1}^{N_*} {x_j b_{j,\lambda} r_\lambda} \Bigg) \otimes G(v_*,\sigma_*)
\end{equation}

\noindent where $b_{j,\lambda}$ is the spectrum of the $j$th SSP normalized at $\lambda_0$, 
$r_\lambda$ is the reddening term, 
$x$ is the population vector whose components $x_j$ (j=1,...,$N_*$) represent the fractional contribution of each SSP to the total synthetic flux at $\lambda_0$,
$M_{\lambda_0}$ is the synthetic flux at the normalization wavelength, 
$G(v_*,\sigma_*)$ is the line-of-sight stellar velocity distribution, and
$\otimes$ denotes the convolution operator. 

Known regions of emission lines in AGNs were masked for the whole sample, whether the galaxy is an AGN or not, to ensure a
homogeneous analysis. 
As only type 2 AGN are present in our sample, the non-stellar AGN contribution
is limited to a few percent of the flux and do not affect the SP parameters derived \citep[Cid-Fernandes, priv. comm.]{cid+04}. Bad pixels or sky background residuals are clipped during the fit, when pixels deviate by more than three times the r.m.s. between
$O_\lambda$ and $M_\lambda$. 
We refer the reader to \citet{cid+05} and references therein for more details on {\sc Starlight}.

The SP models adopted are those of \citet{vazdekis+10} 
with an updated stellar library \citep{miles91}, and cover
the wavelength range 3540\,--\,7400\AA\, at a resolution of FWHM $\sim$ 2.5\AA.
Ages range between 63\,Myr and 18\,Gyr, and metallicities [M/H] between -2.32 and +0.22.
A random fit is shown in Fig.\,\ref{f:fit}, where the observed spectra is shown in black and 
the {\sc Starlight} model in red.

\section{Results}
\label{sec:res}

\begin{figure}
\begin{center}
\epsfig{file=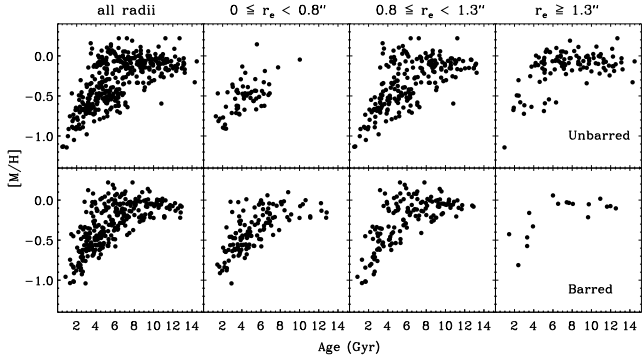,bbllx=40,bblly=70,bburx=594,bbury=380,clip=true,width=\columnwidth}
\caption{Ages vs. metallicities of bulges in unbarred and barred galaxies (top and bottom panels, respectively). Results for the whole sample are shown in the left-hand column, and for different bulge sizes in the remaining columns, as indicated.}
\label{f:stelpop}
\end{center}
\end{figure}

In Fig.\,\ref{f:stelpop} we show the results of our stellar population analysis, where
light-weighted mean ages and metallicites -- normalized at wavelength 4020\AA\, -- are plotted for bulges in barred and unbarred galaxies. 

The robustness of {\sc Starlight} results were thoroughly discussed in \citet{cid+05}, with tests performed also
on SDSS spectra. 
They conclude that measurement uncertainties -- which can be parametrized by S/N -- dominate the final errors. From their Table 1, the uncertainty on the light-weighted parameters are of $\sim$20\% for a S/N\,=\,10 spectrum. Although we have processed all spectra, we rely our analysis solely on the results from spectra with S/N\,$\geq 10$, which constitute 93\% of the sample. 

To circumvent our inability of detecting the short bars in galaxies with Hubble types later Sc, we remove from the analysis all galaxies with bulge-to-total luminosities ratios below 0.043, which is the typical ratio for these latest Hubble types \citep{grawor08}. 

In Fig.\,\ref{f:stelpop} we show the results for the whole sample in the left-hand column, and separated into bulges with $r_e<0.8"$, $0.8"\leq r_e<1.3"$, and $r_e\geq1.3"$ in the remaining columns ($r_e$ in the $i$-band, as given in \citealt{gadotti09}).  
We note that the parameter space covered by the results 
does not change significantly with bulges size,
which is evidence that there are no significant biases from disk contamination inside the SDSS fiber, even for small bulges. Note that bulge light dominates over disk light through $\approx$ 2 times the bulge $r_e$ from the galaxy center, on average \citep[see][]{moriondo98,morelli08,gadotti09}. There seems to be a lack of very old ($>10$ Gyr) bulges in unbarred galaxies with $r_e\leq0.8"$. This is not the case for barred galaxies. If this was a consequence from disk contamination, one would see bulges in unbarred galaxies to have younger ages on average than in barred galaxies (opposite to what we find -- see below). Moreover, stellar populations younger than $\sim 5$ Gyr are seen equally in small and large bulges, barred and unbarred galaxies. It is for these ages that we find a significant difference between bulges in barred and unbarred galaxies. We thus conclude that our results are not affected by disk contamination.

\begin{figure}
\begin{center}
\epsfig{file=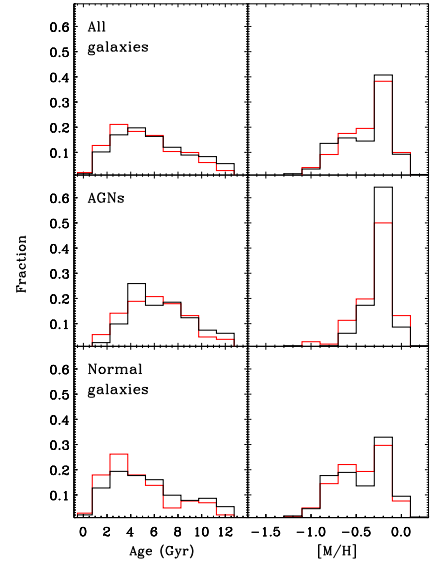,bbllx=30,bblly=230,bburx=400,bbury=697,clip=true,width=0.7\columnwidth}
\caption{Normalized distributions of ages (left-hand column) and metallicities (righ-hand column) for bulges in barred and unbarred galaxies (red and black lines, respectively). Distributions for the whole sample, AGNs and non-active galaxies are given separately. 
This sample has 251 barred and 324 unbarred galaxies (106 and 81 AGNs, respectively).
}
\label{f:distribution0}
\end{center}
\end{figure}

\begin{figure}
\begin{center}
\begin{tabular}{cc}
\epsfig{file=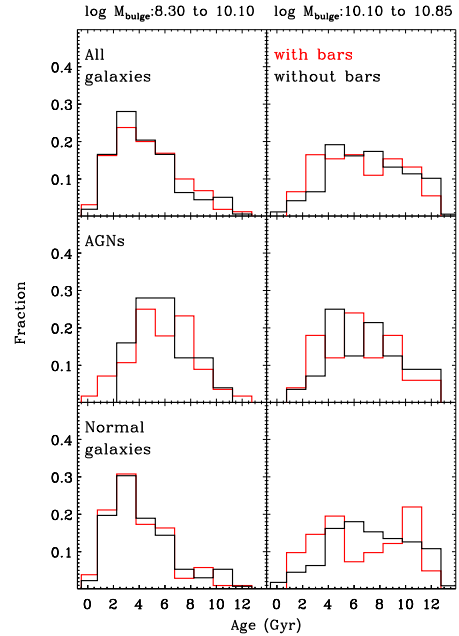,bbllx=35, bblly=230,bburx=375,bbury=715,clip=true,width=0.7\columnwidth}&
\end{tabular}
\caption{Normalized distributions of ages for bulges with $\log M_{bulge} < 10.1$ on the left-hand side, and 
$10.1 < \log M_{bulge} < 10.85$ on the right-hand side.
The low-mass sub-sample has 160 barred and 157 unbarred galaxies (56 and 25 AGNs, respectively) 
and the high-mass sub-sample has 91 barred and 167 unbarred galaxies (50 and 56 AGNs, respectively).
}
\label{f:distribution1}
\end{center}
\end{figure}

In Fig.\,\ref{f:distribution0} we present the normalized distributions of ages and metallicities derived, 
for an upper limit of bulge stellar mass of $10^{10.85}\rm{M}_\odot$ (above which there are no barred galaxies in our sample). 
The top panels show the distributions for the whole sample, and the middle and bottom panels show the distributions
for AGNs and normal galaxies, respectively. 
The age distribution of bulges in barred galaxies shows an excess
of populations younger than $\sim$4\,Gyr. This feature is enhanced when we divide the sample in normal (non-active) galaxies
and AGNs: the excess of young populations is better seen in the distribution of normal galaxies, and disappears 
in AGNs. In fact, a KS test shows that the probabilities that the bulge mean age distributions are drawn from different populations for barred and unbarred galaxies are 98.65\% for the whole sample, 50.13\% for AGNs and 99.94\% for normal galaxies only. Thus, the difference between the mean stellar ages of bulges in non-active barred and unbarred galaxies is significant to a level of almost 4$\sigma$. 

The distributions of {\em bulge} stellar mass for the sample in Fig. \ref{f:distribution0} are, however, statistically different for barred and unbarred galaxies. Barred galaxies have less massive bulges than unbarred galaxies, {\em even though their total stellar mass is similar}, and also have larger disk contamination within the fiber. Therefore, and because less massive bulges tend to have younger mean ages, one cannot tell from Fig. \ref{f:distribution0} alone that bars do indeed turn bulges younger.

For this reason, we inspected several hundred bulge mass intervals in search for those where the distributions of bulge mass, 
and the distributions of disk-to-total light ratios inside the fiber, 
are the same for barred and unbarred galaxies at $\sim1\sigma$ level (ensured with KS tests). We could not find intervals where both AGNs and normal galaxies show equal bulge mass distributions, and then focused on choosing optimal distributions for normal galaxies only (a detailed study of the AGNs will follow in a separate paper). We thus came to two intervals of bulge stellar mass -- below $10^{10.1}\rm{M}_\odot$ and between $10^{10.1}\rm{M}_\odot$ and $10^{10.85}\rm{M}_\odot$ -- whose corresponding results are shown in Fig. \ref{f:distribution1}.

In the lower mass bin, the distributions of bulge ages are statistically similar, in contrast with Fig. \ref{f:distribution0}, which refers to the whole sample. However, for the high-mass bin, the distribution of bulge ages for barred galaxies is clearly bimodal. The mixture modelling statistical KMM test \citep[see][]{kmm} indicates that this is so at a confidence level of 99.9993\%, i.e. $>4\sigma$. The same test results in two normal distributions, with peaks at $4.7$ and 10.4 Gyr. 
We have run the KMM test in all other distributions discussed, and none resulted in a statistical significance larger than $\sim1\sigma$, in particular the distribution of bulge age for unbarred galaxies. This bimodality does not seem to be a result from biased samples, and is a clear evidence of difference between the mean stellar ages in bulges of barred galaxies, as compared to unbarred galaxies.

\begin{figure}
\begin{center}
\epsfig{file=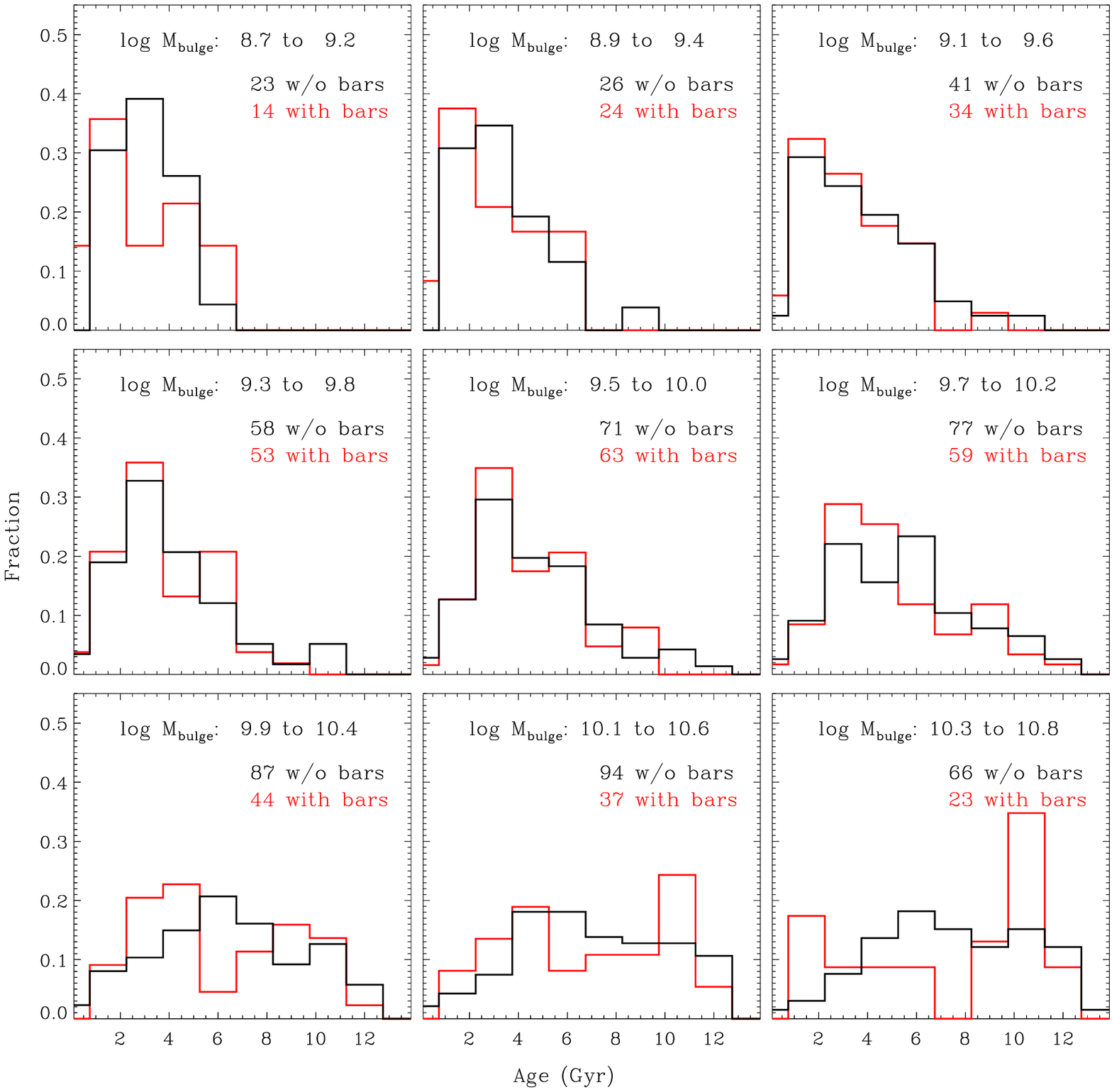,bbllx=0,bblly=0,bburx=620,bbury=620,clip=true,width=\columnwidth}
\caption{Normalized distributions for bulge ages in non-active galaxies, for several mass intervals as indicated. Distributions for barred and unbarred galaxies are shown in red and black lines, respectively.
}
\label{f:bimodality}
\end{center}
\end{figure}

To better inspect this bimodality, we show in Fig. \ref{f:bimodality} the age distributions for several 
bulge mass intervals. 
This is essentially the same as having matched distributions of bulge
mass, since the mass intervals at each panel are narrow.
A first signal of bimodality in the age distribution of non-active galaxies appears in
the interval $\log M_{bulge}$ between 9.7 and 10.2 (the KMM test yields a significance of 99.8928\%) and the peaks reach comparable strengths, as in Fig. \ref{f:distribution1} (significance of 99.9813\%), in the
interval $\log M_{bulge}=$ 9.9 -- 10.4.
A similar
analysis was done for AGNs, corroborating the results above that those effects are inexistent in AGNs.

This result provides statistically-based corroboration that bars have important effects on the processes of bulge building, though the dependence of these effects with bulge mass, and the bimodality in the bulge age distribution, have yet to be explained by theoretical work. 

Furthermore, bars also help building a reservoir of fuel for AGN activity.
{\em In the low bulge mass bin, 35\% of barred galaxies are AGN, whereas this fraction drops to only 16\% when one considers unbarred galaxies. In the high mass bin, 55\% of barred galaxies are AGN, whereas 34\% of unbarred galaxies are AGN.} In both mass bins, if a galaxy is barred it has a higher chance of hosting an AGN. In the high mass bin, processes such as mergers might be acting to help fueling AGN, diminishing the difference in the fraction of AGNs between barred and unbarred galaxies. Bars are not a necessary nor sufficient condition for a galaxy to host an AGN, but these results indicate that in some cases bars do help in fueling AGN activity. There are several studies in the literature -- with opposing results -- on this issue \citep[see e.g.][]{mulreg97,ho97,kna00,lai02}, but they agree that homogeneity on the detection of AGN activity and the assessment on the presence of a bar, as well as sample selection, are critical. Our sample is carefully drawn from a volume-limited sample, 
the data set is homogeneous, and both the AGN and bar classifications are done throughout in a consistent fashion. Nevertheless, we underline that our sample comprises only massive galaxies and that short bars are mostly missed, as discussed above.

Further work is necessary to better understand why the difference between the ages of bulges in barred and unbarred galaxies disappears in AGNs, and why the dichotomy in the bulge ages of massive barred galaxies is not evident in AGNs. A likely interpretation is that feedback from the AGN activity will at some point push gas back from the galaxy center, preventing new star formation episodes \citep[see e.g.][]{schawinski07}.  We stress that even if bulges in barred AGNs are typically less massive, they are {\em not} younger than their unbarred counterparts.

The metallicity distributions show no important
difference between barred and unbarred galaxies, but we will further explore the chemical enrichment in terms of $\alpha$-elements over iron abundances ratios as a function of bulge morphology in a separate paper (P. Coelho \& D. A. Gadotti, in preparation).


We have compared the ages with bar structural parameters 
-- effective surface brightness, effective radius, ellipticity, S\'ersic index, semi-major axis, boxiness and bar-to-total ratio --
but found no evidence for a relation between any of the bar properties and the age of the bulge population.  This is not necessarily surprising, as bulge building by bars depends on complex physical processes and time scales, and the availability of gas. For instance, a bar which was once strong, but is now weakened for any reason [bar weakening is more likely to happen than bar destruction, \citep[e.g.][]{athet05}], could have contributed substantially to build a young population in the bulge of its host galaxy (if enough gas was available), and would now weaken a correlation between bar strength (e.g. ellipticity) and bulge age.

\section{Conclusions}
\label{sec:con}

We derived stellar ages for a sample of 575 bulges in disk galaxies, 251 of those containing bars.
When we consider the sample with bulges stellar masses $<10^{10.85}\rm{M}_\odot$,
we find that the mean stellar ages of bulges of barred galaxies are on average lower than that of unbarred galaxies, at a statistical significance of 99.94\% (or almost 4$\sigma$), when one considers non-active galaxies only. In this sample the galaxy mass distributions are similar between barred and unbarred galaxies.

To make the distributions of {\em bulge} stellar mass of barred and unbarred galaxies similar, we split the sample in two bulge mass bins. We find that bulges in massive non-AGN barred galaxies ($\log M >10^{10.1}\rm{M}_\odot$) show a bimodal stellar age distribution, at a confidence lever of 99.9993\%, or more than $4\sigma$. This can be described as two normal distributions, centered at 4.7 and 10.4\,Gyr. This bimodality  is present above a characteristic mass $\log M_{bulge}=9.7 - 10.2$ and is absent for unbarred galaxies or AGNs. As discussed above, this is a strong observational evidence that corroborates scenarios of bulge building by secular evolution processes induced by bars. On the other hand, this bimodality, and the ages for the two distributions it consists of, are new constraints yet to be explained by successful theories of bar evolution.

We have verified that our results are not caused by biases in the samples compared. We have taken into account, at separate instances, the galaxy mass and bulge mass distributions of barred and unbarred galaxies, in order to have samples with e.g. similar star formation histories. We have also used sub-samples in which the contributions from disk light within the SDSS fiber, through which the spectral information used here is taken, are similarly distributed. Finally, we have not considered very late-type galaxies, for which the presence of a bar cannot be reliably assessed in our sample. Therefore, our results cannot be attributed to a bias in sample selection or a flaw in the methodology we use.

Finally, let us point out again that samples of barred and unbarred galaxies with similar galaxy total stellar mass distributions have statistically significantly different distributions of {\em bulge} mass, in the sense that bulges in barred galaxies have lower masses. This seems to be unexpected because one expects bulge building by bars and, secondly, bars grow from disks and thus one would expect changes in the sense of lower {\em disk} masses. Progress in both theoretical and observational results is needed to clarify this trend.

\section{Acknowledgments}
The authors thank Lia Athanassoula, Roberto Cid-Fernandes, William Schoenell, Lucimara Martins, Rub\'en S\'anchez-Janssen and Alan Alves Brito for useful discussions. We thank the anonymous referee for remarks that greatly improved our paper. PC acknowledges FAPESP support (projects 2008/58406-4 and 2009/09465-0).






\end{document}